\documentclass{SciPost}

\hypersetup{
    colorlinks,
    linkcolor={red!50!black},
    citecolor={blue!50!black},
    urlcolor={blue!80!black}
}

\usepackage[bitstream-charter]{mathdesign}
\DeclareSymbolFont{usualmathcal}{OMS}{cmsy}{m}{n}
\DeclareSymbolFontAlphabet{\mathcal}{usualmathcal}

\fancypagestyle{SPstyle}{
\fancyhf{}
\lhead{\colorbox{scipostblue}{\bf \color{white} ~SciPost Physics Core }}
\rhead{{\bf \color{scipostdeepblue} ~Submission }}

\fancyfoot[C]{\textbf{\thepage}}
}

\begin{document}

\pagestyle{SPstyle}

\begin{center}{\Large \textbf{\color{scipostdeepblue}{
Hierarchic superradiant phases in anisotropic Dicke model\\
}}}\end{center}

\begin{center}\textbf{
D. K. He\textsuperscript{},
Z. Song\textsuperscript{$\star$}
}\end{center}

\begin{center}
School of Physics, Nankai University, Tianjin 300071, China
\\[\baselineskip]
$\star$ \href{mailto:songtc@nankai.edu.cn}{\small songtc@nankai.edu.cn}\,

\end{center}

\section*{\color{scipostdeepblue}{Abstract}}
\textbf{\boldmath{%
We revisit the phase diagram of an anisotropic Dicke model by revealing the
non-analyticity induced by underlying exceptional points. We find that, from a dynamical perspective, the conventional superradiant phase can be further separated into three
regions, in which the systems are characterized by different effective
Hamiltonians, including the harmonic oscillator, the inverted harmonic
oscillator, and their respective counterparts. We employ the Loschmidt echo
to characterize different quantum phases by analyzing the quench dynamics of
a trivial initial state. Numerical simulations for finite systems confirm
our predictions about the existence of hierarchic superradiant phases.
}}

\vspace{\baselineskip}

\noindent\textcolor{white!90!black}{%
\fbox{\parbox{0.975\linewidth}{%
\textcolor{white!40!black}{\begin{tabular}{lr}%
  \begin{minipage}{0.6\textwidth}%
    {\small Copyright attribution to authors. \newline
    This work is a submission to SciPost Physics Core. \newline
    License information to appear upon publication. \newline
    Publication information to appear upon publication.}
  \end{minipage} & \begin{minipage}{0.4\textwidth}
    {\small Received Date \newline Accepted Date \newline Published Date}%
  \end{minipage}
\end{tabular}}
}}
}


\vspace{10pt}
\noindent\rule{\textwidth}{1pt}
\tableofcontents
\noindent\rule{\textwidth}{1pt}
\vspace{10pt}


\section{Introduction}

\label{sec:intro} With the gradual development of experiments on
light-matter interaction \cite%
{stranius2018selective,zhou2019emerging,mueller2020deep,rivera2020light},
the quantum simulation of the Dicke model \cite%
{zhiqiang2017nonequilibrium,marquez2024quantum,black2003observation,baumann2010dicke}
is transitioning from theory to experiment. The Dicke model \cite%
{dicke1954coherence,lambert2004entanglement,emary2003chaos,hepp1973equilibrium,wang1973phase}
is a fundamental model in the field of quantum optics, describing the
interaction between a single-mode light field and $N$ two-level atoms. The
Dicke model has a broad prospect and great potential in the field of quantum
batteries \cite%
{ferraro2018high,andolina2019quantum,crescente2020ultrafast,dou2022extended,quach2022superabsorption,yang2024three,wang2024deep}%
. In the thermodynamic limit ($N\rightarrow \infty $), the ground state of
the Dicke model undergoes a quantum phase transition (QPT) from the normal
phase (NP) to the superradiant phase (SP) \cite%
{emary2003chaos,hepp1973equilibrium,wang1973phase,emary2003quantum,kirton2019introduction,garraway2011dicke,vidal2006finite}
at a certain critical coupling strength, which is referred to as the
superradiant phase transition. In addition to the QPT of the ground state
demonstrated above, the Dicke model also exhibits three distinct phase
transitions, namely, the dissipative phase transition (non-equilibrium
quantum phase transition) \cite%
{bhaseen2012dynamics,kessler2012dissipative,torre2013keldysh}, the
excited-state quantum phase transition \cite%
{kloc2017quantum,kloc2017monodromy,cejnar2021excited}, and the thermal phase
transition \cite{carmichael1973higher,duncan1974effect}.

The concept of exceptional points (EPs) \cite{kato1966,berry2004,heiss2012}, which represents the
degeneracies of non-Hermitian operators, is regarded as a unique feature of
non-Hermitian systems. However, subsequent research has shown that EPs exist
not only in non-Hermitian systems but also in Hermitian systems \cite%
{McDonald_PRX,wang2019non,flynn2020deconstructing,del2022non,wang2022quantum,bilitewski2023manipulating,ughrelidze2024interplay,slim2024optomechanical,busnaina2024quantum,hu2024bosonic,y174-pms8,6lhp-8q6k,he2025hidden}
. The non-analyticity induced by EPs suggests the presence of a phase
transition at this point. In previous studies, we demonstrated that the
superradiant quantum phase transition in the Dicke model can be seen as the
effect of two hidden second-order EPs \cite{he2025hidden,he2025higher}. This quantum phase transition is a dynamical phase transition,
because as the parameter varies, the effective form of the Hamiltonian
changes, resulting in completely different dynamical behaviors on either
side of the transition point. This drives us to seek a more general Dicke
model to investigate its dynamical phase transitions. A more
general version of the Dicke model is called the anisotropic Dicke model 
\cite%
{wang2024deep,buijsman2017nonergodicity,das2023ADMphase,zhu2024quantum,vivek2025self,das2023periodically,chen2024phase}
(ADM), in which the strengths of the rotating-wave and counter-rotating-wave
terms are different. The ADM is being widely studied, including its
applications in quantum batteries \cite{wang2024deep} and the
ergodic-to-nonergodic transition \cite%
{buijsman2017nonergodicity,das2023ADMphase}, as well as work related to
quantum chaos \cite{vivek2025self}.

In this work, we focus on the ADM Hamiltonian and identify the hidden EPs of
this Hamiltonian in the thermodynamic limit. The EPs divide the parameter
space into four regions. The results show that, in addition to the existing
NP to SP transition, there exists a hierarchical structure within the SP
phase. In each region, the original Hamiltonian consists of different
combinations of equivalent Hamiltonians, including the harmonic oscillator
and the inverted harmonic oscillator \cite%
{subramanyan2021physics,barton1986quantum}. The dynamics of such two
oscilators are fundamentally different. Therefore, starting from
an initial state with only a small atomic excitation, the distinct
finite-time dynamical behaviors of the ADM can be used to demonstrate the
existence of EPs and to discriminate between different quantum phases. The finite-time guarantee ensures that the dynamics of an ADM
with a finite atom number can still be accurately described by the
thermodynamic-limit ADM, an idea akin to that proposed in \cite%
{gietka2021inverted}. We employ the Loschmidt echo of quench dynamics to
characterize these phase transitions. The Loschmidt echo can be measured
experimentally using quantum state tomography \cite%
{lvovsky2009continuous,cramer2010efficient,christandl2012reliable}.

The structure of this paper is as follows. In Sec. \ref{Model and
exceptional points}, we introduce the model and pointed out the hidden EPs
within it. In Sec. \ref{Phase diagram}, we solve the Hamiltonian exactly and
present the phase diagram of the model. In Sec. \ref{Quench dynamics}, we
utilize quench dynamics to calculate the Loschmidt echo in order to identify
different dynamical phases. Finally, in Sec. \ref{Summary}, we provide a
summary and discussion. Some details of the calculations are provided in the
Appendix.

\begin{figure}[th]
\centering
\includegraphics[width=1\textwidth]{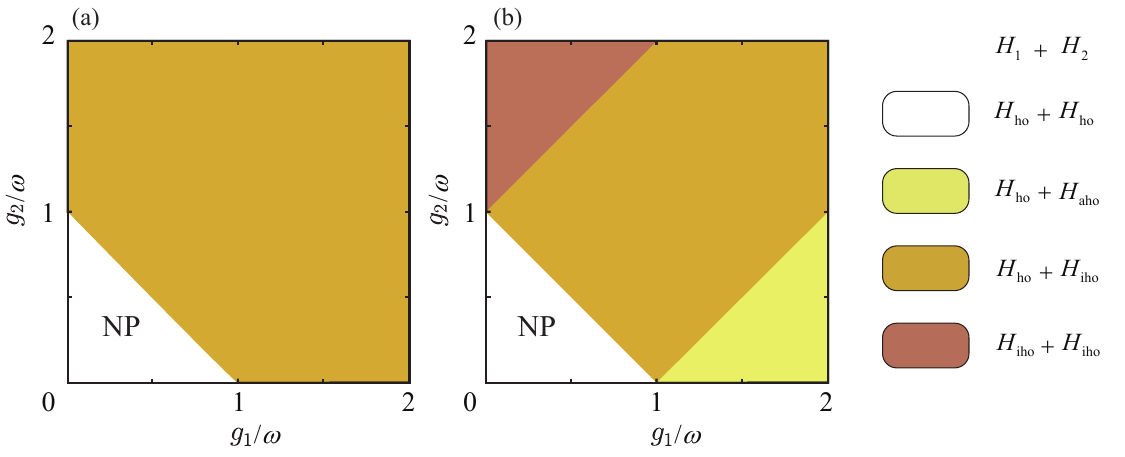}
\caption{Phase diagrams of the Hamiltonian in Eq. (\protect\ref{H_ADM}) on
the parameter $g_{1}g_{2}$ plane, indicating the main conclusion of this
work. Different colors in the diagram distinguish different phases of the
system. (a) The traditional phase diagram of the anisotropic Dicke model
(ADM), obtained by the mean field method, shows that the region $%
g_{1}+g_{2}<\omega$ corresponds to the normal phase (NP), and the region $%
g_{1}+g_{2}>\omega$ corresponds to the superradiant phase (SP). (b) The phase
diagram of the ADM, revealed by the underlying exceptional points (EPs) of
the effective Hamiltonian in Eq. (\protect\ref{Heff})\ of the system, shows
that the original superradiant phase (a) can be further divided into three
distinct phases. We label these phases as $\text{SP}_{1}$, $\text{SP}_{2}$,
and $\text{SP}_{3}$, respectively. The corresponding equivalent Hamiltonians
of the effective Hamiltonian in each region are indicated in the panel.
Here, we assume $\protect\omega =\protect\omega _{0}$.}
\label{fig1}
\end{figure}

\section{Model and exceptional points}

\label{Model and exceptional points} We consider a Hamiltonian of a
single-mode boson coupled to $N$ two-level atoms, where the rotating-wave
and counter-rotating-wave terms are distinct. This model is known as the ADM.%
\begin{eqnarray}
H &=&\omega a^{\dag }a+\omega _{0}J_{z}+\frac{g_{1}}{\sqrt{N}}\left( a^{\dag
}J_{-}+aJ_{+}\right)  \nonumber \\
&&+\frac{g_{2}}{\sqrt{N}}\left( a^{\dag }J_{+}+aJ_{-}\right) .  \label{H_ADM}
\end{eqnarray}%
Here, $a^{\dag }$ and $a$ represent the creation and annihilation operators
of the single-mode boson, respectively. $J_{\pm }$ and $J_{z}$ are the
collective atomic operators, and their commutation relations are as follows%
\begin{equation}
\left[ a,a^{\dag }\right] =1,\left[ J_{z},J_{\pm }\right] =\pm J_{\pm }, %
\left[ J_{+},J_{-}\right] =2J_{z}.
\end{equation}%
The first and second terms of the Hamiltonian represent the free
Hamiltonians of the light field and the $N$ two-level atoms, respectively,
with their strengths controlled by $\omega $ and $\omega _{0}$. The third
and fourth terms correspond to the rotating-wave and counter-rotating-wave
coupling terms, with coupling strengths $g_{1}$ and $g_{2}$, respectively.
When $g_{1}=g_{2}$, the model reduces to the Dicke model. For convenience,
in the following derivations, we assume $\omega =\omega _{0}$, $g_{1}>0$, $%
g_{2}>0$. The phase diagram of the ADM has been conclusively established in
previous studies based on the mean field method \cite{das2023ADMphase}. In
the parameter plane of $g_{1} g_{2}$, the region where $g_{1}+g_{2}>\omega $
corresponds to the superradiant phase, while the region where $%
g_{1}+g_{2}<\omega $ corresponds to the normal phase. {Although the standard Dicke model in cavity QED is prohibited from exhibiting the superradiant phase due to the no-go theorem\cite{vukics2012adequacy,nataf2010no}, recent studies have shown that anisotropy can overcome the no-go theorem\cite{chen2024phase,ye2025superradiant}, thus providing a theoretical basis for its potential experimental realization in platforms such as cavity QED.} The phase diagram is
shown in Fig. \ref{fig1}(a).

In the following, we will show that the conventional superradiant phase can
be further separated into three regions, in which the systems are
characterized by different effective Hamiltonians in large $N$ limit,
including the harmonic oscillator, the inverted harmonic oscillator, and
their respective counterparts. We refer to these as hierarchic superradiant
phases because the same given initial state exhibits distinct dynamic
behaviors.

We introduce the Holstein-Primakoff (HP) transformation to convert the spin
operators into bosonic operators $b$%
\begin{eqnarray}
J_{z} &=&b^{\dag }b-\frac{N}{2},  \nonumber \\
J_{+} &=&\left( J_{-}\right) ^{\dag }=b^{\dag }\sqrt{N-b^{\dag }b},
\end{eqnarray}%
In the thermodynamic limit where $N\rightarrow \infty $ and neglecting
constant terms, the Hamiltonian can be rewritten as%
\begin{eqnarray}
H_{\mathrm{eff}} &=&\omega \left( a^{\dag }a+b^{\dag }b\right) +g_{1}\left(
a^{\dag }b+ab^{\dag }\right)  \nonumber \\
&&+g_{2}\left( a^{\dag }b^{\dag }+ab\right) .  \label{Heff}
\end{eqnarray}%
$H_{\mathrm{eff}}$ can be regarded as a two-site Hermitian bosonic Kitaev
model \cite{he2025hidden,he2025higher,vidal2007entanglement}. In previous studies, we revealed
that this model possesses hidden EPs. We introduce a linear transformation 
\begin{equation}
d_{1,2}=\frac{1}{\sqrt{2}}(a \pm b),
\end{equation}%
 to decompose $H_{\mathrm{eff}}$ into two independent subspaces Hamiltonian
can be written as
\begin{eqnarray}
H_{\mathrm{eff}} &=&H_{1}+H_{2}  \nonumber \\
&=&\phi _{L}\left( 
\begin{array}{cc}
h_{1} & 0 \\ 
0 & h_{2}%
\end{array}
\right) \phi _{R}.
\end{eqnarray}%
The non-Hermitian Nambu spinor is defined as $\phi _{L}=\left(
d_{1},-d_{1}^{\dag },d_{2},-d_{2}^{\dag }\right) $ and $\phi _{R}=\left(
d_{1}^{\dag },d_{1},d_{2}^{\dag },d_{2}\right) ^{T}$. {This representation has been studied in Ref. \cite{McDonald_PRX,wang2019non,flynn2020deconstructing}, and it can be generalized to arbitrary quadratic bosonic systems.} The forms of the two
matrices are
\begin{equation}
h_{1,2}=\frac{1}{2}\left( \omega \pm g_{1}\right) \sigma _{z}\pm \frac{i}{2}
g_{2}\sigma _{y},
\end{equation}
$h_{1,2}$ are non-Hermitian matrices, and $\sigma _{z}$ and $\sigma _{y}$
are Pauli matrices, defined as%
\begin{equation}
\sigma _{z}=\left( 
\begin{array}{cc}
1 & 0 \\ 
0 & -1%
\end{array}
\right) ,\sigma _{y}=\left( 
\begin{array}{cc}
0 & -i \\ 
i & 0%
\end{array}
\right) .
\end{equation}%
{The eigenvalues of $h_{1,2}$ are}
\begin{eqnarray}
\lambda _{1}^{\pm } &=&\pm \frac{1}{2}\sqrt{\left( \omega +g_{1}\right)
^{2}-g_{2}^{2}},  \nonumber \\
\lambda _{2}^{\pm } &=&\pm \frac{1}{2}\sqrt{\left( \omega -g_{1}\right)
^{2}-g_{2}^{2}}.
\end{eqnarray}%
{The corresponding right eigenvectors are}
\begin{eqnarray}
\phi _{1}^{\pm } &=&\left( 
\begin{array}{c}
-\frac{1}{g_{2}}\left( \omega +g_{1}+2\lambda _{1}^{\pm }\right) \\ 
1%
\end{array}
\right) ,  \nonumber \\
\phi _{2}^{\pm } &=&\left( 
\begin{array}{c}
\frac{1}{g_{2}}\left( \omega -g_{1}+2\lambda _{2}^{\pm }\right) \\ 
1%
\end{array}
\right) .
\end{eqnarray}%
From the forms of the eigenvalues and eigenvectors, we can see that the
matrices possess EPs. $h_{1}$ has a second-order EP when$\ \left\vert \omega
+g_{1}\right\vert =\left\vert g_{2}\right\vert $, and $h_{2}$ has a
second-order EP when$\ \left\vert \omega -g_{1}\right\vert =\left\vert
g_{2}\right\vert $. These EP can divide different regions in the $%
g_{1}-g_{2} $ parameter plane, as shown in Fig. \ref{fig1}(b). In the next
section, we will provide the exact solutions for the diagonalized
Hamiltonian in each region.

\section{Phase diagram}

\label{Phase diagram} The Hamiltonians $H_{1}$ and $H_{2}$ can be explicitly
expressed as follows:

\begin{equation}
H_{1}=\left( \omega +g_{1}\right) d_{1}^{\dag }d_{1}+\frac{g_{2}}{2}\left(
d_{1}^{\dag }d_{1}^{\dag }+d_{1}d_{1}\right) ,
\end{equation}%
and%
\begin{equation}
H_{2}=\left( \omega -g_{1}\right) d_{2}^{\dag }d_{2}-\frac{g_{2}}{2}\left(
d_{2}^{\dag }d_{2}^{\dag }+d_{2}d_{2}\right) ,
\end{equation}%
respectively. We note that the two Hamiltonians have the same form as%
\begin{equation}
\mathcal{H}=\mu \beta ^{\dag }\beta +\frac{\Delta }{2}\left( \beta ^{\dag
}\beta ^{\dag }+\beta \beta \right) ,
\end{equation}%
where $\beta $ is the bosonic annihilation operator. In the Appendix \ref%
{Appendix_A}, we provide the derivation of the diagonalization of the
Hamiltonian $\mathcal{H} $, based on which two Hamiltonians $H_{1}$\ and\ $%
H_{2}$\ can be reduced to different simple form in the four regions in the
first quadrant of $g_{1}g_{2}$ plane.

Ignoring the energy constants, there exist three types of equivalent
Hamiltonians, given by

\begin{eqnarray}
H_{\text{\textrm{ho}}} &=&\Omega _{i}\left( \gamma _{i}^{\dag }\gamma _{i}+%
\frac{1}{2}\right) ,  \label{ho} \\
H_{\text{\textrm{iho}}} &=&\left( -1\right) ^{i+1}\frac{\Omega _{i}}{2}\left[
\left( \gamma _{i}^{\dag }\right) ^{2}+\gamma _{i}^{2}\right] ,  \label{iho}
\\
H_{\text{\textrm{aho}}} &=&-\Omega _{i}\left( \gamma _{i}^{\dag }\gamma _{i}+%
\frac{1}{2}\right) ,  \label{aho}
\end{eqnarray}%
with $i=1$, $2$, where $\gamma _{i}$\ are bosonic annihilation operators.
The positive factor $\Omega _{i}$ is given by
{\color{red}
\begin{eqnarray}
\Omega _{1} &=&\omega\left|\sqrt{\left( 1 + \frac{g_{1}}{\omega}\right) ^{2}-\left(\frac{g_{2}}{\omega}\right)^{2}}\right|, \\
\Omega _{2} &=&\omega\left|\sqrt{\left( 1 - \frac{g_{1}}{\omega}\right) ^{2}-\left(\frac{g_{2}}{\omega}\right)^{2}}\right|.
\end{eqnarray}}
The harmonic oscillator Hamiltonian $H_{\text{\textrm{ho}}}$ is the standard
form of the Hamiltonian for a harmonic oscillator. {The inverted harmonic
oscillator Hamiltonian $H_{\text{\textrm{iho}}}$ describes a system with an
inverted potential; its eigenenergies are continuous and unbounded\cite{barton1986quantum},
rendering the system unstable and allowing it to tunnel toward states with
higher particle numbers\cite{scully1997quantum}.} In Appendix \ref{Appendix_B}, we present a detailed
account of the dynamical characteristics of this Hamiltonian. The
anti-harmonic oscillator Hamiltonian $H_{\text{ \textrm{aho}}}$ is the
negative of the standard harmonic oscillator Hamiltonian. {For $H_{2}$, when $|\omega-g_{1}|>g_2$ and $\omega-g_{1}<0$, the system diagonalizes into such an anti-harmonic oscillator. The anti-harmonic oscillator describes a system where the vacuum state has the highest energy, and states with higher particle numbers have lower energies. Its dynamics under isolated conditions are oscillatory, just like those of a standard harmonic oscillator.} Each of these Hamiltonians has distinct physical properties
and implications for the stability and behavior of the system. 
{Under the dynamics of an isolated system, the harmonic oscillator and anti-harmonic oscillator are stable, while the inverted harmonic oscillator is unstable.} In the following, we present the explicit form of the equivalent Hamiltonians in each region.

\begin{figure*}[th]
\centering
\includegraphics[width=1\textwidth]{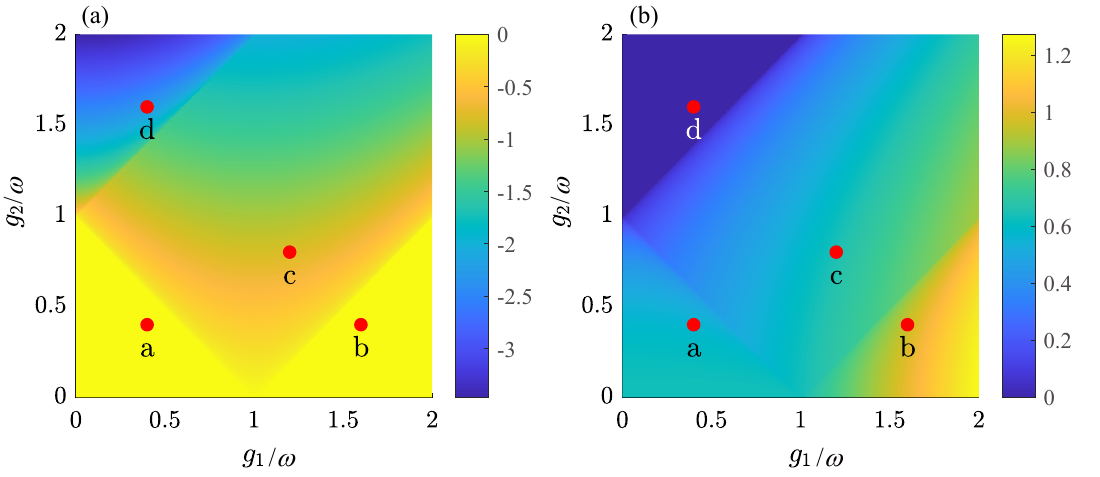}
\caption{The plots of the decay rate $\protect\lambda $ in (a), given by Eq.
(\protect\ref{decay_rate}) and frequency $f$ in (b), given by Eq. (\protect
\ref{frequency}) of the effective Hamiltonian {\color{red}on the $g_{1}/\omega$-$g_{2}/\omega$ plane.} It
can be seen from the figures that there are clear distinctions between
different phases in terms of $\protect\lambda $ and $f$ . Four
representitive points in each regions are selected, indicated by red dots at
the same positions in both panels, with coordinates a$(0.4,0.4)$, b$%
(1.6,0.4) $, c$(1.2,0.8)$, and d$(0.4,1.6)$. The corresponding quench
dynamical behaviors of the original ADM in finite systems at these points,
obtained by numerical simulations, are presented in Fig. \protect\ref{fig3}.}
\label{fig2}
\end{figure*}

(i) For $g_{1}+g_{2}<\omega $, in this region, the two Hamiltonians have the
form

\begin{equation}
H_{1}=\Omega _{1}\left( \gamma _{1}^{\dag }\gamma _{1}+\frac{1}{2}\right) - 
\frac{1}{2}\left( \omega +g_{1}\right) ,
\end{equation}%
and%
\begin{equation}
H_{2}=\Omega _{2}\left( \gamma _{2}^{\dag }\gamma _{2}+\frac{1}{2}\right) - 
\frac{1}{2}\left( \omega -g_{1}\right) ,
\end{equation}%
respectively. Here, $\gamma _{1}$\ and $\gamma _{2}$\ are bosonic
annihilation operators, given by%
\begin{equation}
\gamma _{i}=\sinh \left(\theta _{i}\right)d_{i}^{\dag }+\cosh \left(\theta _{i}\right)d_{i},
\label{gamma i}
\end{equation}%
with%
\begin{equation}
\tanh \left(\theta _{1}\right)=\frac{\omega +g_{1}-\Omega _{1}}{g_{2}},  \label{tanh0_1}
\end{equation}%
and%
\begin{equation}
\tanh \left(\theta _{2}\right)=\frac{\omega -g_{1}-\Omega _{2}}{g_{2}},  \label{tanh0_2}
\end{equation}%
respectively. {This part corresponds to the $\mathrm{NP}$ region in Fig. \ref{fig1}(b).}

(ii) For $g_{1}+g_{2}>\omega $ and $g_{2}<g_{1}-\omega $,\ in this region,
two Hamiltonians have the form%
\begin{equation}
H_{1}=\Omega _{1}\left( \gamma _{1}^{\dag }\gamma _{1}+\frac{1}{2}\right) - 
\frac{1}{2}\left( \omega +g_{1}\right) ,
\end{equation}%
and%
\begin{equation}
H_{2}=-\Omega _{2}\left( \gamma _{2}^{\dag }\gamma _{2}+\frac{1}{2}\right) - 
\frac{1}{2}\left( \omega -g_{1}\right) ,
\end{equation}%
respectively. {Here, $\gamma _{1}$, $\gamma _{2}$, $\tanh\left(\theta _{1}\right)$ and $\tanh\left(\theta _{2}\right)$ have the same forms in
Eqs. (\ref{gamma i}),(\ref{tanh0_1}) and (\ref{tanh0_2}). This part corresponds to the $\mathrm{SP_1}$ region in Fig. \ref{fig1}(b).}

(iii) For $g_{1}+g_{2}>\omega $ and $g_{1}-\omega <g_{2}<g_{1}+\omega $, in
this region, two Hamiltonians have the form%
\begin{equation}
H_{1}=\Omega _{1}\left( \gamma _{1}^{\dag }\gamma _{1}+\frac{1}{2}\right) - 
\frac{1}{2}\left( \omega +g_{1}\right) ,
\end{equation}%
and%
\begin{equation}
H_{2}=i\frac{\Omega _{2}}{2}\left[ \left( \gamma _{2}^{\dag }\right)
^{2}+\left( \gamma _{2}\right) ^{2}\right] -\frac{1}{2}\left( \omega
-g_{1}\right) ,
\end{equation}
respectively. Here, $\gamma _{1}$\ and $\gamma _{2}$\ have the same forms in
Eq. (\ref{gamma i}), but with%
\begin{equation}
\tanh \left(\theta _{1}\right)=\frac{\left( \omega +g_{1}\right) -\Omega _{1}}{g_{2}},
\label{tanh2_1}
\end{equation}%
and%
\begin{equation}
\tanh \left(\theta _{2}\right)=\frac{g_{2}-i\Omega _{2}}{\omega -g_{1}}.  \label{tanh2_2}
\end{equation}
{This part corresponds to the $\mathrm{SP_2}$ region in Fig. \ref{fig1}(b).}

(iv) For $g_{1}+g_{2}>\omega $ and $g_{1}+\omega <g_{2}$, in this region,
two Hamiltonians have the form%
\begin{equation}
H_{1}=-i\frac{\Omega _{1}}{2}\left[ \left( \gamma _{2}^{\dag }\right)
^{2}+\left( \gamma _{2}\right) ^{2}\right] -\frac{1}{2}\left( \omega
+g_{1}\right) ,
\end{equation}%
and%
\begin{equation}
H_{2}=i\frac{\Omega _{2}}{2}\left[ \left( \gamma _{2}^{\dag }\right)
^{2}+\left( \gamma _{2}\right) ^{2}\right] -\frac{1}{2}\left( \omega
-g_{1}\right) ,
\end{equation}%
respectively. Here, $\gamma _{1}$\ and $\gamma _{2}$\ have the same forms in
Eq. (\ref{gamma i}), but with%
\begin{equation}
\tanh \left(\theta _{1}\right)=\frac{g_{2}+i\Omega _{1}}{\omega +g_{1}},  \label{tanh3_1}
\end{equation}%
and%
\begin{equation}
\tanh \left(\theta _{2}\right)=\frac{g_{2}-i\Omega _{2}}{\omega -g_{1}},  \label{tanh3_2}
\end{equation}%
respectively. {This part corresponds to the $\mathrm{SP_3}$ region in Fig. \ref{fig1}(b).} The corresponding equivalent Hamiltonians are indicated in the
phase diagram shown in Fig. \ref{fig1}(b). It shows that the configurations
of the equivalent Hamiltonians are different in each region. The whole
superradiant phase is separated three sub-phases, which are refered to as
hierarchic superradiant phases. Here, we would like to emphasize that the
phase diagram presented here is not a zero-temperature phase diagram.
Different equivalent Hamiltonians exhibit different dynamics, which cannot
be captured by mean-field theory. These phases have to be detected by the
measurement of information in the excited state. Building upon this insight,
we will propose a dynamic demonstration of the phase diagram.

\section{Quench dynamics}

\label{Quench dynamics} 
\begin{figure*}[h]
\centering
\includegraphics[width=0.8\textwidth]{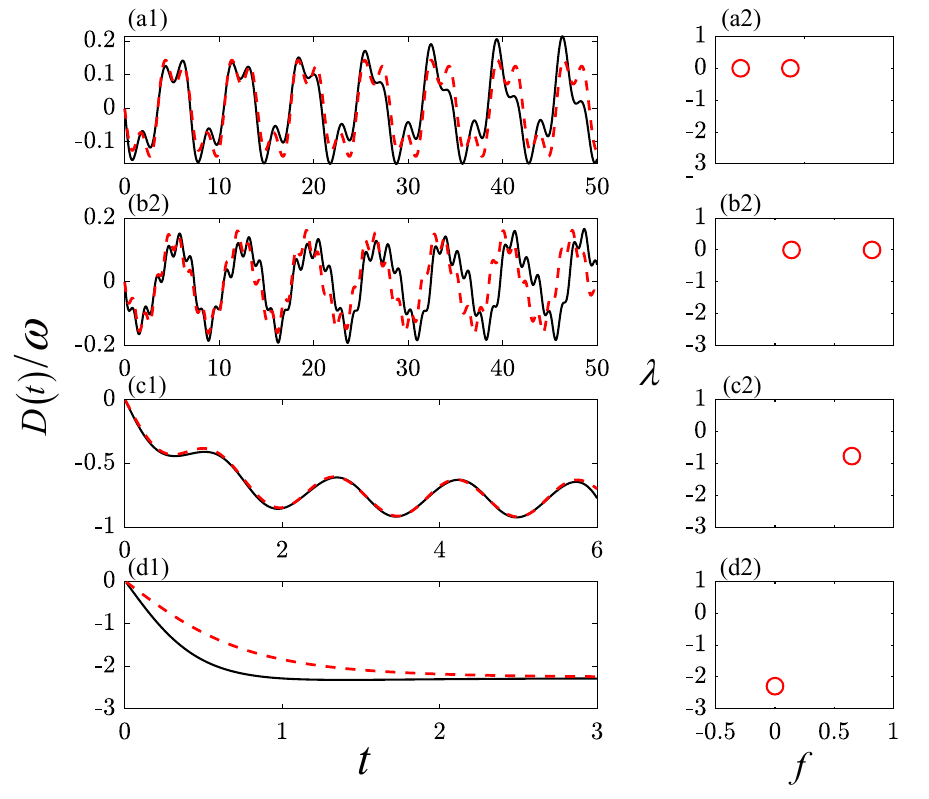}
\caption{The plots of $D(t)$, given by Eq. (\protect\ref{D_t}), and their
characteristics for the original ADM, given by Eq. (\protect\ref{H_ADM})  and effective Hamiltonian $H_{\text{eff}}$ given by Eq. (\protect
\ref{Heff}) in finite systems at the represented points indicated in Fig. 
\protect\ref{fig2}. The plots in (a1)-(d1) are obtained by numerical
simulations, the solid black line represents the numerical
results obtained from the full ADM, whereas the red dashed line corresponds
to the analytical result obtained from Eq. (\protect\ref{Lo}). The
corresponding decay rates $\protect\lambda $\ and frequencies $f$, plotted
in (a2)-(d2), are extracted from the plots of {\color{red}$D(t)/\omega$}. The number of atoms in
the system is $N=100$, and the bosonic Hilbert space is truncated at $n_{\text{max}}=140$. {Employing a larger bosonic cutoff $n_{\text{max}}$
  or increasing the number of atoms $N$ does not alter the system's dynamical behavior over any finite time interval.} These results are in accordance with the predictions from the analysis of the effective Hamiltonians. {\color{red}The time in the figure is in units of $\omega^{-1}$}}
\label{fig3}
\end{figure*}
In this section, we investigate the dynamic behavior of the phase diagram,
including the hierarchical superradiant phases. We consider the quench
dynamics under the postquench Hamiltonian $H$. We conduct numerical
simulations for the Loschmidt echo, defined as

\begin{equation}
L\left( t\right) =\left\vert \left\langle \psi \left( 0\right) |\psi \left(
t\right) \right\rangle \right\vert ^{2},
\end{equation}%
which is a measure of the revival for the initial state $\left\vert \psi
\left( 0\right) \right\rangle $. It allows us to characterize the properties
of a system, provided that a proper initial state is chosen. We choose the
empty state as the initial state $\left\vert \psi \left( 0\right)
\right\rangle =\left\vert \Downarrow \right\rangle \left\vert 0\right\rangle 
$ and calculate its evolved state%
\begin{equation}
\left\vert \psi \left( t\right) \right\rangle =\exp \left( -iHt\right)
\left\vert \psi \left( 0\right) \right\rangle ,
\end{equation}%
where states $\left\vert \Downarrow \right\rangle $ and $\left\vert
0\right\rangle $ are defined by $J_{z}\left\vert \Downarrow \right\rangle
=-N/2\left\vert \Downarrow \right\rangle $ and $a\left\vert 0\right\rangle
=0 $, respectively. Before the computation for the finite ADM system, we
would like to estimate the possible result.

We start with the investigation for the effective Hamiltonian $H_{\mathrm{eff%
}}$, which can be dealt with analytically. The corresponding initial state
becomes $\left\vert \psi \left( 0\right) \right\rangle =\left\vert
0\right\rangle _{a}\left\vert 0\right\rangle _{b}$\ and evolved state is\ $%
\left\vert \psi \left( t\right) \right\rangle =\exp \left( -iH_{\mathrm{eff}%
}t\right) \left\vert \psi \left( 0\right) \right\rangle $, correspondingly. 
Note that the initial state can also be written in the form $
\left\vert \psi \left( 0\right) \right\rangle =\left\vert 0\right\rangle
_{d_{1}}\left\vert 0\right\rangle _{d_{2}}$, satisfying $d_{1}\left\vert
\psi \left( 0\right) \right\rangle =d_{2}\left\vert \psi \left( 0\right)
\right\rangle =0$, which allows the product form of $L\left( t\right) $. {In the thermodynamic limit,} the
Loschmidt echo has the following approximate expressions in each regions  
\begin{equation}
L\left( t\right) \approx \left\{ 
\begin{array}{lc}
\left[ 1-2A^{2}\sin ^{2}\left( \Omega _{1}t\right) \right] \left[
1-2B^{2}\sin ^{2}\left( \Omega _{2}t\right) \right] , & \mathrm{NP} \\ 
\left[ 1-2A^{2}\sin ^{2}\left( \Omega _{1}t\right) \right] \left[
1-2B^{2}\sin ^{2}\left( \Omega _{2}t\right) \right] , & \mathrm{SP}_{1} \\ 
\left[ 1-2A^{2}\sin ^{2}\left( \Omega _{1}t\right) \right] \left[ \cosh
\left( \Omega _{2}t\right) \right] ^{-1}, & \mathrm{SP}_{2} \\ 
\left[ \cosh \left( \Omega _{1}t\right) \right] ^{-1}\left[ \cosh \left(
\Omega _{2}t\right) \right] ^{-1}, & \mathrm{SP}_{3}%
\end{array}%
\right. ,  \label{Lo}
\end{equation}%
where the parameters $A$ and $B$ are given explicitly as
\begin{equation}
A=\frac{2}{1+2\tanh ^{-2}\theta _{1}}=\frac{2(\omega +g_1-\mathrm{sgn}(\omega+g_1)\Omega_{1})}{(\omega+g_1-\Omega_1)^2+2g_2^2},\label{A}
\end{equation}
and
\begin{equation}
B=\frac{2}{1+2\tanh ^{-2}\theta _{2}}=\frac{2(\omega -g_1-\mathrm{sgn}(\omega-g_1)\Omega_{1})}{(\omega-g_1-\Omega_2)^2+2g_2^2}, \label{B}
\end{equation}
respectively. The details of the calculation can be found in Appendix
 \ref{Appendix_B}. {In each region,} $L\left(t\right) $ is the product of two functions, which take different configurations. For the $\mathrm{SP}_{1}$ region, it is the product of two
periodic functions. For the $\mathrm{SP}_{2}$ region, it is the product of a
periodic function and a decaying function. For the $\mathrm{\ SP}_{3}$
region, it is the product of two decaying functions. The reason
why the product of the two functions adopts distinct configurations in
different regions is that the underlying Hamiltonians are combined
differently: whenever the effective Hamiltonian contains a harmonic or
anti-harmonic oscillator, it supplies the periodic factor, whereas the
presence of an inverted harmonic oscillator provides the decaying factor.
We refer these phases to as hierarchic superradiant phases. {It is noteworthy that the phase structure revealed by the Loschmidt echo exhibits considerable robustness with respect to the choice of initial states. We further investigate an intriguing special case where the system is initially prepared with all atoms in the excited state and the optical field in the vacuum state. Through a global SU(2) spin rotation transformation $J_z \rightarrow -J_z$, $J_\pm \rightarrow J_\mp$, the dynamical behavior under this initial condition can be rigorously mapped to the case with the ground state as the initial condition\cite{bilitewski2023manipulating,duha2024two,duha2025nonequilibrium}, which constitutes the main focus of this work. This symmetry operation leads to an important physical consequence: the effective phase diagram measured from this initial state becomes a mirror image of the phase diagram shown in Fig. \ref{fig1}(b) specifically manifesting as an exchange between the $\mathrm{SP}_1$ and $\mathrm{SP}_3$ regions, while the $\mathrm{SP}_2$ region remains unchanged.}

We note that the function $\left[ \cosh \left( \Omega _{i}t\right) \right]
^{-1}\approx 2e^{-\Omega _{i}t}$, decaying exponentially with rate $\Omega
_{i}$, after long time scale. Then, the oscillating frequency and the decay
rate can be the dynamic characters of the hierarchic SPs. In order to
characterize the hierarchy of the phases, we focus on the quantity%
\begin{equation}
D(t)=\frac{\partial }{\partial t}\ln L\left( t\right) ,  \label{D_t}
\end{equation}%
because we have%
\begin{equation}
\frac{\partial }{\partial t}\ln e^{-\Omega _{i}t}=-\Omega _{i}.
\end{equation}%
It is expected that $D(t)$\ is the sum of two simple functions, which take
different configurations\ in each region of superradiant phases. Therefore,
the factors $\Omega _{1}$\ and $\Omega _{2}$ can be extracted from the
long-time behavior of $D(t)$. For the \textrm{SP}$_{1}$ region, $D(t)$
oscillates around zero, from which two frequencies $f_{1}=\Omega _{1}/\pi $\
and $f_{2}=\Omega _{2}/\pi $ can be extracted. In the \textrm{SP}$_{2}$
region, it oscillates around a constant, from which the oscillating
frequency $f_{1}$ and the balance point $-\Omega _{2}$\ can be extracted. In
the \textrm{SP}$_{3}$ region, it decays to a constant, from which the decay
rate $\lambda =-\left( \Omega _{1}+\Omega _{2}\right) $\ can be extracted.
What is shown in Fig. \ref{fig2} is the analytical result of the decay rate {\color{red}
\begin{equation}
	\lambda =-\left( \Omega _{1}+\Omega _{2}\right)/\omega ,  \label{decay_rate}
\end{equation}}
and the sum of frequencies {\color{red}
\begin{equation}
	f=\left(f_{1}+f_{2}\right)/\omega,  \label{frequency}
\end{equation}}
which can be extracted from the echo of the evolved state of the effective
Hamiltonian $H_{\mathrm{eff}}$. We can see the non-analytical behaviors of
the plots at the phase boundaries.

Now, we turn to the computation of the corresponding quantities for the
original ADM Hamiltonian. For a system with a finite number of atoms, the
dimension of the Hilbert space is infinite. Therefore, the time evolution of
the initial state is computed using exact diagonalization under the
truncation approximation. The computations are performed using a uniform
mesh in the time discretization for the {truncated} matrix. We selected four
representative points in the four phases of the ADM to perform quench
dynamics verification, and the results are shown in Fig. \ref{fig3}. {Within the time scales of our numerical simulations, our results do not depend on the matrix size.} The
extracted decay rate $\lambda $ and frequency $f$ correspond to those in
Fig. \ref{fig2}. The results are in accordance with the predictions from the
analysis of the effective Hamiltonians. This demonstrates that there indeed
exist hierarchical superradiant phases within the traditional superradiant
phase of the ADM.

\section{Summary}

\label{Summary}

In summary, we have demonstrated that the conventional superradiant phase
can be further separated into three regions. The underlying mechanism is the
existence of the exceptional points in the effective Hamiltonians in the
thermodynamic limit. {Unlike traditional quantum phase transitions that typically occur in the ground state of the system, this constitutes a dynamical phase transition where the phase separations arise from sudden changes in the complete set of eigenstates.} In this sense,
the proposed phase diagram is not merely a mathematical concept, but
definitely results in evident observations. Numerical simulations have been
performed to compute the Loschmidt echo for finite systems. The results
indicate that such observables are sufficient to characterize the
hierarchical superradiant phases.

\section*{Acknowledgements}

\paragraph{Funding information}

This work was supported by the National Natural Science Foundation of China
(under Grant No. 12374461).

\begin{appendix}

	\section{Diagonalization of the Hamiltonians}
	
	\label{Appendix_A}
\numberwithin{equation}{section}

\label{Appendix} \setcounter{equation}{0} \renewcommand{\theequation}{A%
	\arabic{equation}}

In this appendix, we provide the derivation of the diagonalization of the
Hamiltonian $\mathcal{H}$, which is equivalent to the two Hamiltonians $%
H_{1} $ and $H_{2}$\ given in the main text. The Hamiltonian reads%
\begin{equation}
	\mathcal{H}=\mu \beta ^{\dag }\beta +\frac{\Delta }{2}\left( \beta ^{\dag
	}\beta ^{\dag }+\beta \beta \right) ,
\end{equation}%
where $\beta $ is the bosonic annihilation operator. Here, we do not
restrict the range of $\mu $ and $\Delta $, and $\mathcal{H}$ naturally
satisfies%
\begin{eqnarray}
	H_{1} &=&\mathcal{H}\left( \mu =\omega +g_{1},\Delta =g_{2}\right) ,  \notag
	\\
	H_{2} &=&\mathcal{H}\left( \mu =\omega -g_{1},\Delta =g_{2}\right) .
\end{eqnarray}%
We assume that there exists a Bogoliubov transformation 
\begin{equation}
	\gamma =\sinh \left(\theta\right) \beta ^{\dag }+\cosh \left(\theta\right) \beta ,
\end{equation}%
that allows for the diagonalization of the Hamiltonian $\mathcal{H}$. Here, $%
\gamma $ is also the bosonic annihilation operator and the inverse
transformation is%
\begin{equation}
	\beta =\cosh \left(\theta\right) \gamma -\sinh \left(\theta\right) \gamma ^{\dag }.
\end{equation}%
The coefficient $\theta $\ is determined by the following process.
Substituting the transformation into $\mathcal{H}$ we have%
\begin{eqnarray}
	\mathcal{H} &=&\frac{1}{2}\left[ \Delta \cosh \left(2\theta\right) -\mu \sinh \left(2\theta\right)\right] \left[ \left( \gamma ^{\dag }\right) ^{2}+\gamma ^{2}\right]  \notag
	\\
	&&+\left[ \mu \cosh ^{2}\left(\theta\right) -\frac{\Delta }{2}\sinh \left(2\theta\right) \right]
	\gamma ^{\dag }\gamma  \notag \\
	&&+\left[ \mu \sinh ^{2}\left(\theta\right) -\frac{\Delta }{2}\sinh \left(2\theta\right) \right]
	\left( 1+\gamma ^{\dag }\gamma \right) .
\end{eqnarray}%
We consider the following two cases respectively.

(i) $\left\vert \mu \right\vert >\left\vert \Delta \right\vert $, the
Hamiltonian can be written as the diagonalized form 
\begin{equation}
	\mathcal{H}=\text{sgn}(\mu )[\sqrt{\mu ^{2}-\Delta ^{2}}\left( \gamma ^{\dag
	}\gamma +\frac{1}{2}\right) ]-\frac{\mu }{2},
\end{equation}%
when we take 
\begin{equation}
	\tanh \left(\theta\right) =\frac{\mu -\text{sgn}(\mu )\sqrt{\mu ^{2}-\Delta ^{2}}}{\Delta }. \label{tanh_1}
\end{equation}%
(ii) $\left\vert \mu \right\vert <\left\vert \Delta \right\vert $, the
Hamiltonian can be written as the anti-diagonalized form%
\begin{equation}
	\mathcal{H}=\text{sgn}(\Delta )\frac{1}{2}\{\sqrt{\Delta ^{2}-\mu ^{2}}%
	[\left( \gamma ^{\dag }\right) ^{2}+\gamma ^{2}]\}-\frac{\mu }{2},
\end{equation}%
when we take%
\begin{equation}
	\tanh \left(\theta\right) =\frac{\Delta -\text{sgn}(\Delta )\sqrt{\Delta ^{2}-\mu ^{2}}}{\mu }. \label{tanh_2}
\end{equation}
\numberwithin{equation}{section}

	\section{Calculation of the Loschmidt echos}
	\label{Appendix_B}
	In this appendix, we present the derivations of the evolved states $\left|\psi\left(t\right)\right>$, given in Eqs. (\ref{psit_ho_aho}) and (\ref{psit_iho}), respectively, and the corresponding Loschmidt echo $L_{j}$, given in Eq. (\ref{L_t}), respectively, for the initial state $\left|0\right>_{d_{j}}$, which is the vacuum state of the operator $d_{j}$ (where $j=1,2$).  The driven Hamiltonians are $H_\text{ho}$, $H_\text{aho}$, and $H_\text{iho}$, given in Eqs. (\ref{ho}), (\ref{iho}) and (\ref{aho}), respectively.
	
	The initial state $\left|0\right>_{d_{j}}$ can be spanned by the common eigenstates $\left\{\left|l\right>_{j},l\in \left[0,\infty\right)\right\}$ of the Hamiltonians $H_{\mathrm{ho}}$, $H_{\mathrm{aho}}$ and $H_{\mathrm{iho}}$, in the form
	\begin{equation}
	\left\vert 0\right\rangle _{d_{j}}=\sum_{l=0}^{\infty }\left[ \tanh (\theta_{j})\right] ^{l}A_{l}\left\vert 2l\right\rangle _{j},
	\end{equation}
	where $\left\vert l\right\rangle _{j}=\frac{1}{\sqrt{l!}}\left( \gamma_{j}^{\dag }\right) ^{l}\left\vert 0\right\rangle _{\gamma _{j}}$ and $\left\vert 0\right\rangle _{\gamma _{j}}$ is the vacuum state of the
	operator $\gamma _{j}$. The coefficients $\tanh \theta _{j}$  is defined in Eqs. (\ref{tanh_1}) and (\ref{tanh_2}) , and given explicitly in the main text in Eqs. (\ref{tanh0_1}), (\ref{tanh0_2}), (\ref{tanh2_1}), (\ref{tanh2_2}), (\ref{tanh3_1}) and (\ref{tanh3_2}). $A_{l}$ obey the iteration relation $A_{l+1}\sqrt{2l+2}$ $=A_{l}\sqrt{2l+1}$, where $A_{0}$ is a constant determined by normalization. Hence, the evolved
	states $\left\vert \psi (t)\right\rangle _{j}$\ for the Hamiltonians $H_{\mathrm{ho}}$ and $H_{\mathrm{aho}}$\ can be directly obtained as
	\begin{equation}
	\left\vert \psi (t)\right\rangle _{j}=\left\{ 
	\begin{array}{c}
	\exp \left( -iH_{\mathrm{ho}}t\right) \left\vert 0\right\rangle
	_{d_{j}}=\sum_{l=0}^{\infty }\exp \left( -i2l\Omega _{j}t\right) \left[
	\tanh (\theta _{j})\right] ^{l}A_{l}\left\vert 2l\right\rangle _{d_{j}}, \\ 
	\exp \left( -iH_{\mathrm{aho}}t\right) \left\vert 0\right\rangle
	_{d_{j}}=\sum_{l=0}^{\infty }\exp \left( i2l\Omega _{j}t\right) \left[ \tanh
	(\theta _{j})\right] ^{l}A_{l}\left\vert 2l\right\rangle _{d_{j}},%
	\end{array}%
	\right. .\label{psit_ho_aho}
	\end{equation}
	However, the set of states $\left\{ \left\vert l\right\rangle _{j}\right\} $ are no longer the eigenstates of $H_{\mathrm{iho}}$. We have to take	another approach to derive the corresponding $\left\vert \psi(t)\right\rangle _{j}$.\ We note that the time evolution operator $U(t)=\exp\left( -iH_{\mathrm{iho}}t\right) $\ is nothing but the squeezing operator	in quantum optics \cite{subramanyan2021physics,scully1997quantum}. This allows us to establish the relation
	\begin{equation}
	\mathcal{H}(t)U(t)\left\vert 0\right\rangle _{\gamma_{j}}=\mathcal{H}(t)\left\vert \psi (t)\right\rangle _{j}=0,
	\end{equation}
	with 
	\begin{equation}
	\mathcal{H}(t)=\gamma _{j}\cosh \left( \Omega _{j}t\right) +\gamma
	_{j}^{\dag }i(-1)^{j+1}\sinh \left( \Omega _{j}t\right) .
	\end{equation}
	This indicates that the evolved state $\left\vert \psi (t)\right\rangle _{j}$ is the instantaneous zero-energy eigenstate of the auxiliary	time-dependent Hamiltonian $\mathcal{H}(t)$. The evolved state can be obtained as
	\begin{equation}
	U(t)\left\vert 0\right\rangle _{\gamma_{j}} =\sqrt{\mathrm{sech}\left( \Omega
	_{j}t\right) }\sum_{l=0}^{\infty }\frac{\sqrt{\left( 2l\right) !}}{l!2^{l}}%
	\left[ (-1)^{j}i\tanh( \left\vert \Omega_j \right\vert t)\right] ^{l}\left\vert
	2l\right\rangle _{\gamma_{j}},
	\end{equation}
	by using the series method. If the initial state is the vacuum state $\left|0\right>_{d_{j}}$ of the operator $d_{j}$, then the evolved state can be expressed as
	\begin{equation}
	\left|\psi(t)\right>_{j}=\sum_{l=0}^{\infty}\frac{1}{\sqrt{2l!}}\mathcal{H}(t)^{2l}\left[\tanh(\theta_{j})\right]^{l}A_{l}U(t)\left|0\right>_{\gamma_j}.\label{psit_iho}
	\end{equation}
	Then the {corresponding} $L_{j}\left( t\right)$are obtained as 
	\begin{equation}
	L_{j}\left( t\right) \approx \left\{ 
	\begin{array}{c}
	\left\vert \left\langle 0\right\vert _{d_{j}}\exp \left( -iH_{\mathrm{ho}%
	}t\right) \left\vert 0\right\rangle _{d_{j}}\right\vert ^{2}=1-\frac{8\tanh
	^{2}\left( \theta _{j}\right) }{\left[ 2+\tanh ^{2}\left( \theta _{j}\right) %
	\right] ^{2}}\sin ^{2}\left( \Omega _{j}t\right) , \\ 
	\left\vert \left\langle 0\right\vert _{d_{j}}\exp \left( -iH_{\mathrm{iho}%
	}t\right) \left\vert 0\right\rangle _{d_{j}}\right\vert ^{2}=\left[ \cosh
	\left( \Omega _{j}t\right) \right] ^{-1}, \\ 
	\left\vert \left\langle 0\right\vert _{d_{j}}\exp \left( -iH_{\mathrm{aho}%
	}t\right) \left\vert 0\right\rangle _{d_{j}}\right\vert ^{2}=1-\frac{8\tanh
	^{2}\left( \theta _{j}\right) }{\left[ 2+\tanh ^{2}\left( \theta _{j}\right) %
	\right] ^{2}}\sin ^{2}\left( \Omega _{j}t\right) ,%
	\end{array}
	\right. . \label{L_t}
	\end{equation}
	{This corresponds to Eqs. (\ref{Lo}), (\ref{A}), and (\ref{B}) in the main text.}

\end{appendix}





\bibliographystyle{plain}
\bibliography{SciPost_Example_BiBTeX_File.bib}


\end{document}